\documentclass[sigplan,10pt,screen]{acmart}

\AtBeginDocument{%
	\providecommand\BibTeX{{%
			\normalfont B\kern-0.5em{\scshape i\kern-0.25em b}\kern-0.8em\TeX}}}

\setcopyright{acmcopyright}
\copyrightyear{2022}
\acmYear{2022}
\setcopyright{rightsretained}
\acmConference[EuroSys '22]{Seventeenth European Conference on
	Computer Systems}{April 5--8, 2022}{RENNES, France}
\acmBooktitle{Seventeenth European Conference on Computer Systems
	(EuroSys '22), April 5--8, 2022, RENNES,
	France}
\acmDOI{10.1145/3492321.3519565}
\acmISBN{978-1-4503-9162-7/22/04}

\usepackage{wrapfig}
\usepackage{comment}
\usepackage{colortbl}

\usepackage{hyperref}

\usepackage{amsthm} 

\usepackage{color,soul}
\usepackage{graphics}
\usepackage{hyperref}
\usepackage{lipsum}
\usepackage{tikz}
\usepackage{enumitem}	
\usepackage{listings} 
\usepackage{booktabs}	
\usepackage{lipsum}

\usepackage{capt-of}	

\usepackage{todonotes}  
\usepackage{subcaption} 

\definecolor{refkey}{rgb}{249,158,26}
\definecolor{labelkey}{rgb}{0,1,0}

\usepackage{mathrsfs}	
\usepackage{amsfonts}

\usepackage[export]{adjustbox} 

\usepackage{algorithm,algorithmic,amsmath} 
\usepackage{boldline}					

\usepackage{multirow}

\renewcommand{\paragraph}[1]{\vskip 3pt\noindent\textbf{#1 }}	 

  {\begin{list}{$\bullet$}%
     {\setlength{\parsep}{0pt}%
      \setlength{\topsep}{0pt}%
      \setlength{\itemsep}{2pt}}}%
  {\end{list}}
\newcommand\Note[1]{\sethlcolor{red} \hl{#1}} 
\newcommand\Noted[1]{} 

\newcommand\xzlNote[1]{\sethlcolor{yellow} \hl{#1}} 

\definecolor{darkblue}{rgb}{0.0, 0.0, 0.55}
\definecolor{mygreen}{HTML}{ADFF2F}
\definecolor{mylightgray}{gray}{0.8}

\newenvironment{myitemize}%
  {\begin{itemize}
	[leftmargin=0cm,
		itemindent=.3cm,
		labelwidth=\itemindent,
		labelsep=0pt,
		parsep=1pt,
		topsep=1pt,
		itemsep=1pt,
		align=left]
  }%
  {\end{itemize}}    

\newenvironment{myenumerate}%
  {\begin{enumerate}
	[leftmargin=.cm,itemindent=.5cm,labelwidth=\itemindent,
		labelsep=0pt,
		parsep=1pt,
		topsep=1pt,
		itemsep=3pt,
		align=left]
  }%
  {\end{enumerate}}    

\newcommand\sect[1]{Section~\ref{sec:#1}}	

\newcommand{\code}[1]{\texttt{\small{#1}}}	

\newcommand{\sys}{our recorder}

\newcommand{\teedriver}{driverlet}
\newcommand{\Teedriver}{Driverlet}
\newcommand{\teedrivers}{driverlets}

\newcommand{\template}{interaction template}
\newcommand{\Template}{Interaction template}

\newcommand{\TZ}{TrustZone}

\newcommand{\rpi}{RPi3}
\newcommand{\arm}{Arm}

\makeatletter
\def\@copyrightspace{\relax}
\makeatother

\begin{document}

\title{Minimum Viable Device Drivers for ARM TrustZone}
 

\author{Liwei Guo}
\affiliation{%
	\institution{University of Virginia}
	\city{}
	\state{}
	\country{}
}
\email{lg8sp@virginia.edu}

\author{Felix Xiaozhu Lin}
\affiliation{%
	\institution{University of Virginia}
	\city{}
	\state{}
	\country{}
}
\email{felixlin@virginia.edu}


\date{}


\begin{abstract}
While TrustZone can isolate IO hardware, 
it lacks drivers for modern IO devices. 
Rather than porting drivers, 
we propose a novel approach to deriving minimum viable drivers: 
developers exercise a full driver and record the driver/device interactions; 
the processed recordings, dubbed \textit{driverlets}, are replayed in the TEE at run time to access IO devices. 
\Teedriver{}s address two key challenges: correctness and expressiveness, 
for which they build on a key construct called \textit{interaction template}. 
The interaction template ensures faithful reproduction of recorded IO jobs (albeit on new IO data); 
it accepts dynamic input values; 
it tolerates nondeterministic device behaviors. 
We demonstrate \teedriver{}s on a series of sophisticated devices, making them accessible to TrustZone for the first time to our knowledge.
Our experiments show that \teedriver{}s are secure, easy to build, and incur acceptable overhead (1.4$\times$-2.7$\times$ compared to native drivers). 
\Teedriver{}s fill a critical gap in the TrustZone TEE, 
realizing its long-promised vision of secure IO. 



\end{abstract}



\maketitle


\begin{figure}[t]
	\centering
	\includegraphics[width=0.48\textwidth{}]{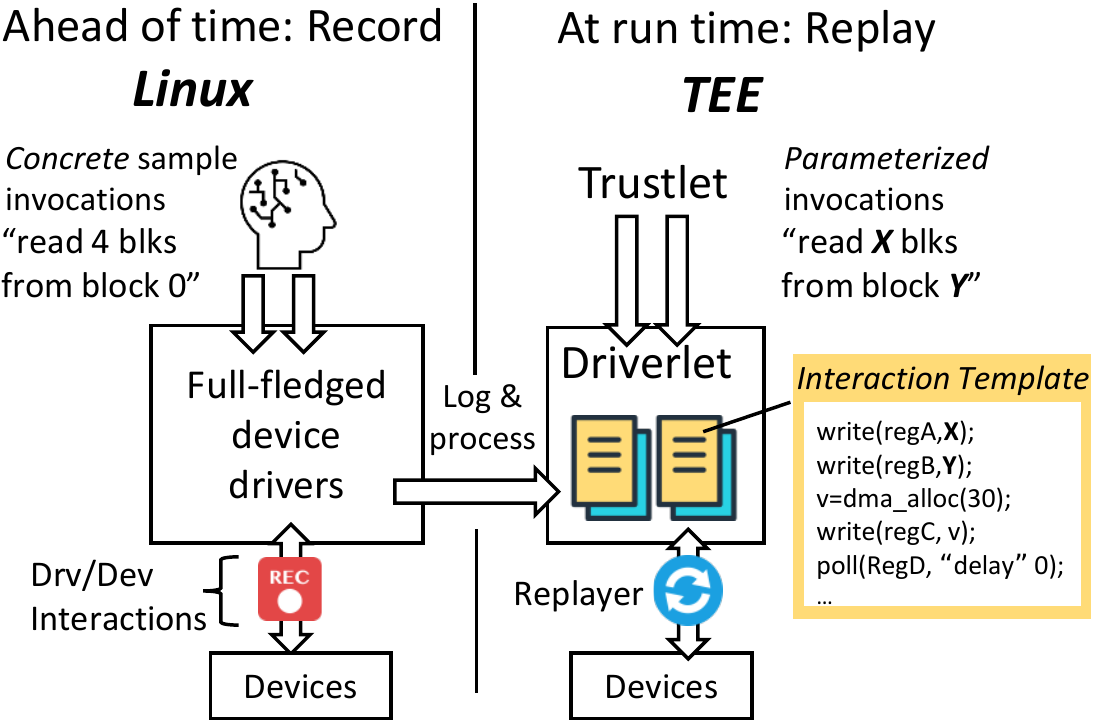}
	\caption{Overview. Our system records driver/device interactions from full-fledged device drivers and automatically processes them into interaction templates. The results are lightweight driverlets for serving secure IO in TEE.}
	\label{fig:workflow}
\end{figure}

\section{Introduction}
\label{sec:intro}

Arm TrustZone is a trusted execution environment (TEE). 
It hosts small programs called trustlets, which manage sensitive data against an untrusted OS. 
Designed for IO-rich client devices, TrustZone features \textit{secure IO}: 
the TEE maps an IO device's physical resources -- registers, interrupts, and memory regions -- to the TEE's address space, therefore keeping 
them inaccessible to the OS. 
Trustlets can use secure IO for: 
(1) storing credentials, keys, and biometric data~\cite{android-trustlets,tz-healthdata};
(2) acquiring sensitive audio and video for processing~\cite{tinystack,ppfl,darknetz}; 
(3) rendering graphical UI with security-critical contents~\cite{rushmore}. 
In these use cases, 
IO data moves between trustlets and IO devices, bypassing the OS; 
the trusted computing base (TCB) consists of only the TEE and the underlying hardware. 


Secure IO, however, remains largely untapped today. 
Popular TrustZone runtimes such as OPTEE and Trusty have been developed for almost one decade~\cite{trusty-git, optee-010};
yet they still cannot access important IO -- flash storage, cameras, and display controllers~\cite{optee-changelog}.
The difficulty lies in device drivers. 
Implementing drivers -- even only supporting a small set of functions needed by trustlets -- can be non-trivial. 
For instance, to read a block from an SD card, a multimedia card (MMC) driver issues more than 1000 register accesses: configuring the MMC controller, exchanging 5--6 commands/responses with the storage hardware, and orchestrating DMA transfers. 
To do it correctly, developers need to reason about the bus protocols from over 200 pages of MMC standards~\cite{sd-standard,emmc-standard} and the device's register interface. They also need to deal with hardware quirks or bugs~\cite{rpi3-hw-bug}.

How about reusing mature drivers from a commodity OS? 
That route is difficult as well. 
Mature drivers are designed to be comprehensive. 
They are often large, structured as multiple abstraction layers. 
They depend on a variety of kernel services. 
For instance, the MMC driver in Linux comprises 25K--30K SLoC scattered in over 150 files~\cite{transkernel}; 
it invokes kernel APIs such as the slab allocator, DMA, and CPU scheduling. 
Developers could port the driver in two ways. 
(1) 
They move a driver and all its dependencies to the TEE (i.e. ``lift and shift''). 
This, however, is likely to move excessive, or even most, the Linux kernel code, adding hundreds of K SLoC~\cite{sgx-lkl}. 
It violates TrustZone's principle of a lean TCB. 
(2) Alternatively, 
developers may refactor the driver, stripping code unneeded by the trustlets. 
To do so, they nevertheless 
have to reason about the driver and device internals and port a variety of kernel APIs. 
Our experiences in \sect{case} show high efforts of reasoning, debugging, and trial-and-error. 

\paragraph{Approach}
Since complex device drivers are overkills to simple trustlets,
we advocate for a new way for deriving the drivers: 
instead of reusing a mature driver's \textit{code},
we selectively reuse its \textit{interactions} with the device. 
The basic idea is shown in Figure~\ref{fig:workflow}. 
(1) 
Developers exercise a mature driver with sample invocations that would be made by the trustlets, 
e.g. write 10 blocks at block address 42.  
(2) 
With symbolic tracing, our recorder logs driver/device interaction events: register accesses, shared memory accesses, and interrupts. 
(3)
At run time, as the trustlets invoke the driver interfaces, the TEE replays the recorded interactions. 
At a higher level, our approach resembles \textit{duck typing} in the context of a device driver: upon user inputs, a driverlet executes as a driver; 
upon device inputs, a driverlet reacts as a driver; 
then it must be as correct as a driver.

Compared to driver porting, our approach requires less developer effort. 
Developers only reason about the driver/device \textit{interfaces} for recording, while remaining oblivious to their complex \textit{internals}. 
Our approach respects TEE's security needs. 
The record runs are done on a developer's machine, an environment considered trustworthy~\cite{supplychain-whitepaper}.
The resultant recordings comprise only primitive events but no complex code. 
The replayer is as simple as a few KSLoC and imposes stringent security checks.  
\sect{eval} will present a detailed security analysis. 

\paragraph{Challenges}
(1) \textit{Correctness}. 
By following pre-recorded interactions, 
how does the replayer assure that it faithfully reproduces the recorded IO jobs, e.g. having written given data to block address 42?
This is exacerbated by the device's nondeterministic behaviors, e.g.
it may return different register values or different interrupts in response to the same driver stimuli. 
(2) \textit{Expressiveness}. 
While the recorder exercises the driver 
with a finite set of concrete inputs, 
e.g. \{\textit{blkid=42, blkcnt=4}\}, 
can the recorded interactions be generalized to cover larger \textit{regions} of the input space, e.g. \{\textit{0<blkid<0xffff, 0<blkcnt<32}\}?



\paragraph{Driverlet}
We present a system for recording and replaying driver/device interactions. 
For a given IO device, the recorder produces multiple recordings, dubbed a \textit{driverlet}.
The driverlet offers satisfactory coverage of IO functions as needed by the trustlets, e.g. to access flash storage at a variety of block granularities. 
A driverlet thus serves as the device's minimum viable driver.

A \teedriver{} provides the same level of correctness guarantee as the full driver with the following insight: 
a replay run is \textit{faithful} when the device makes the same \textit{state transitions} as in the record run. 
Based on this insight, 
the recorder identifies a series of state-changing events from the recorded interactions. 
The state-changing events are the ``waypoints'' that a faithful replay must precisely reproduce. 
As for other interaction events not affecting the device state, 
the recorder emits constraints on them to tolerate their deviations from the recording in a principled manner. 
This approach sets us apart from many record-and-replay systems~\cite{r2,tinystack,knockoff},
where the replays must reproduce the recorded executions with high precision 
including all the non-deterministic events. 
From each record run, 
the recorder distills an \textit{\template{}}, 
which, at a high level, specifies the behavior envelope for the replay run. 
The template prescribes \textit{input events} that the replayer should expect, 
which may come from the trustlet program, the device, or the TEE environment; the template also prescribes \textit{output events} that the replayer should emit to the latter. 
By reproducing the sequence of input/output events, 
the replayer is guaranteed to induce the same device state transitions as in the record run, hence reproducing the IO jobs faithfully.

An \template{} is more expressive than a raw log of interactions in three ways. 
(1) Input events accept dynamic values that are not limited to the recorded concrete values. 
(2) Output events can be parameterized, encoding the data dependency between the earlier inputs and the later outputs.
(3) 
A driver's polling loops are lifted as meta events, each replayed as a varying number of input/output events until the loop termination condition is met. 

\paragraph{Results}
We implement our system with a suite of known techniques: taint tracking, selective symbolic execution, and static code analysis. 
We apply our approach to a variety of device drivers previously considered too complex to TEE: MMC, USB mass storage, and CSI cameras. 
With light developer efforts and no knowledge of device internals (e.g. the device FSM specifications), our recorder generates \teedrivers{}, each comprising 3-10 \template{}s and 50-1500 interaction events. 
The replayer itself has minimum dependencies on the TEE; 
it is in around 1000 SLoC, which are three orders of magnitude smaller than the full drivers. 
The \teedrivers{} incur modest overheads:
on RaspberryPi 3, a low-cost Arm platform, 
trustlets can execute 100 SQLite queries per second (1.4$\times$ slower than a full-fledged native driver) and 
capture images at 2.1 FPS from a CSI camera (2.7$\times$ slower). 
They provide practical performance to use cases such as secure storage and surveillance.

\paragraph{Contributions}
This paper contributes: 

\begin{myitemize}
	\item 
	A new model called the \textit{\teedriver{}} for reusing driver/device interactions via record and replay.
	The \teedriver{}s generalize the recordings as 	interaction templates, 
	ensuring sound replay while accepting new inputs beyond those being recorded. 
	
	\item 
	A toolkit for automatically exercising a full driver and generating  \teedrivers{}. 

	\item 
	Case studies of applying \teedriver{}s to a variety of complex device drivers.
	The resultant \teedriver{}s are immediately deployable to TrustZone. With them, the TrustZone TEE gains access to these devices for the first time to our knowledge.
			
\end{myitemize}


By fixing a long missed link in TrustZone, 
\teedriver{}s enable holistic, end-to-end protection of the TrustZone's IO.


\section{Motivations}
\label{sec:bkgnd}

\subsection{Example trustlets of secure IO}
The following use cases motivate our design.
In these cases, TEE protects the IO, 
which prevents the OS from observing sensitive IO data and tampering with the data. 

\begin{myitemize}
	\item Secure storage. 
	Trustlets manage sensitive data, such as credentials, fingerprints, and user emails. 
	The trustlets store and retrieve the data by using an in-TEE flash hardware such as multimedia cards (MMC)~\cite{sbd} and embedded MMC (eMMC)~\cite{strongbox}. 
	
	\item Trusted perception. 
	Trustlets ingest sensitive data from the sensor hardware protected within the TEE. 
	A particularly interesting case is video, which often contains privacy-sensitive contents and requires non-trivial camera drivers~\cite{tz-image-processing}. 
	
	\item Trusted UI. 
	Trustlets render security-sensitive contents, such as cloud verification codes and bank account information; 
	the UI reads in user inputs such as key presses and touches. 
	Both the display controller and input devices are isolated in TEE~\cite{rushmore,schrondintext}. 
	
\end{myitemize}

\paragraph{IO needed by today's trustlets}
The above use cases all rely on trustworthy device drivers, 
for which we exploit the following opportunities. 

\begin{myitemize}
	\item 
	\textit{Trustlets are less sensitive to IO performance.}
	Today's trustlets are mostly deployed on mobile devices and tolerate overhead~\cite{android-trustlets}.
	For instance, those managing user credentials or email contents do not need high storage throughput; 
	trustlets for surveillance can be served with median to low frame rates and resolutions (e.g. 720P at 1FPS~\cite{diva,focus}). 

	\item 
	\textit{Trustlets can be served with simple IO device features.}
	In the examples above, trustlets only need the write/read of flash blocks, acquisition of video frames, and rendering bitmaps or vector paths to given screen coordinates. 
	They are unlikely to need complex device features, e.g. hotplug of flash cards. 
	
	\item 
	\textit{Trustlets may share IO devices at coarse-grained time intervals.} 
	Concurrent trustlets are far fewer than the normal-world apps. 
	Even when multiple trustlets request to access the same device, their requests can be serialized without notable user experience degradation. 
	Examples include serializing accesses to a credential storage and serializing drawing requests to a trusted display. 
	
	
\end{myitemize}

The above observations on existing trustlets motivate our design choice:  trading IO performance and fine-grained sharing for driver simplicity and security. 
As future trustlets may be more performance-demanding, our approach may fall short; we will discuss the limitation in Section~\ref{sec:bkgnd:limit}.



%



\subsection{Prior art}
\label{sec:design:prior}

\paragraph{Mature device drivers}
They are often overkills to trustlets because of their following design choices.
(1) \textit{Optimal performance}. 
For instance, an MMC driver tunes bus parameters periodically (by default every second). 
It implements a complex state machine for handling corner cases, so that the driver can recover from runtime errors with minimum loss of work. 
%
(2) \textit{Device dynamism}. 
The drivers support full device features as permitted by hardware specifications, e.g. runtime power management~\cite{device-pm} and device hotplug~\cite{usb-hotplug}.
(3) \textit{Fine-grained sharing}. 
A driver maintains separate contexts for concurrent apps. 
The driver multiplexes requests at short time intervals. 
Some drivers, e.g. USB, implement sophisticated channel scheduling for optimizing throughput and meeting request deadlines. 
\paragraph{Existing approaches}
The following approaches may be used to bring drivers into TEE.
\begin{myenumerate}
	\item 
	\textit{Lift and shift.} 
	One may bundle a full driver with all its kernel dependencies and port them to the TEE~\cite{sgx-lkl,rumpkernel}. 
	While easing porting, 
	this approach often adds substantial code (tens of KSLoc) as well as bringing vulnerabilities to the TEE. 
	
	\item 
	\textit{Trim down.}
	Developers may carve out only needed driver functions from the kernel, which has been done on simpler drivers, e.g. UART~\cite{optee-uart-driver}.
	On non-trivial drivers, however, the approach is impractical because the drivers have deep dependencies on complex device configurations and kernel frameworks. 
	\sect{eval} presents our own experiences on trimming three drivers. 
	
	\item 
	\textit{Synthesis.} 
	While it is possible to synthesize some drivers from scratch (assuming the device FSM is fully known), 
	this approach requires non-trivial efforts. 
	Developers have to write FSM specifications~\cite{termite}, develop code templates~\cite{termite-2}, and write glue code. 
	In case of secure IO, we deem such synthesis efforts unwarranted, 
	because trustlets only need simple IO functions and the source code of a mature driver is already available. 
	
	\item 
	\textit{Partitioning.}
	One may partition a full driver, executing only the security-sensitive partition in the secure world and leaving the remainder in the normal world~\cite{trustshadow,trustui}.
	This approach, however faces two obstacles. 
	First, it is built upon the \textit{Trim down} approach, where the developer must manually carve out code pieces and resolve kernel dependencies.
	Second, it is difficult for the developer to reason about the security properties at the interface between secure/insecure partitions, where notorious attacks are common~\cite{iago-attack}.
\end{myenumerate}

\section{Approach overview}

\subsection{System model}
\label{sec:bkgnd:model}


\paragraph{SoC hardware}
We assume an IO device \textit{instance} exclusively assigned to the TrustZone TEE. 
This is feasible as a modern SoC often has multiple independent instances of a device, e.g. 4--6 MMC controllers~\cite{imx6sx,imx7,tegra2rtm}. 
Through an address space controller (TZASC), the SoC maps selected physical memory and device registers to the TEE. 

\paragraph{Targeted IO devices}  
We focus on IO devices that have strong use cases in TEE. 
These devices manipulate sensitive IO data while lacking end-to-end data encryption. 
Examples include USB storage, 
video/audio devices, 
and display controllers. 
We make the following assumptions. 

%
%
%
%
%
%

\begin{figure}
	\centering
	\includegraphics[width=0.48\textwidth{}]{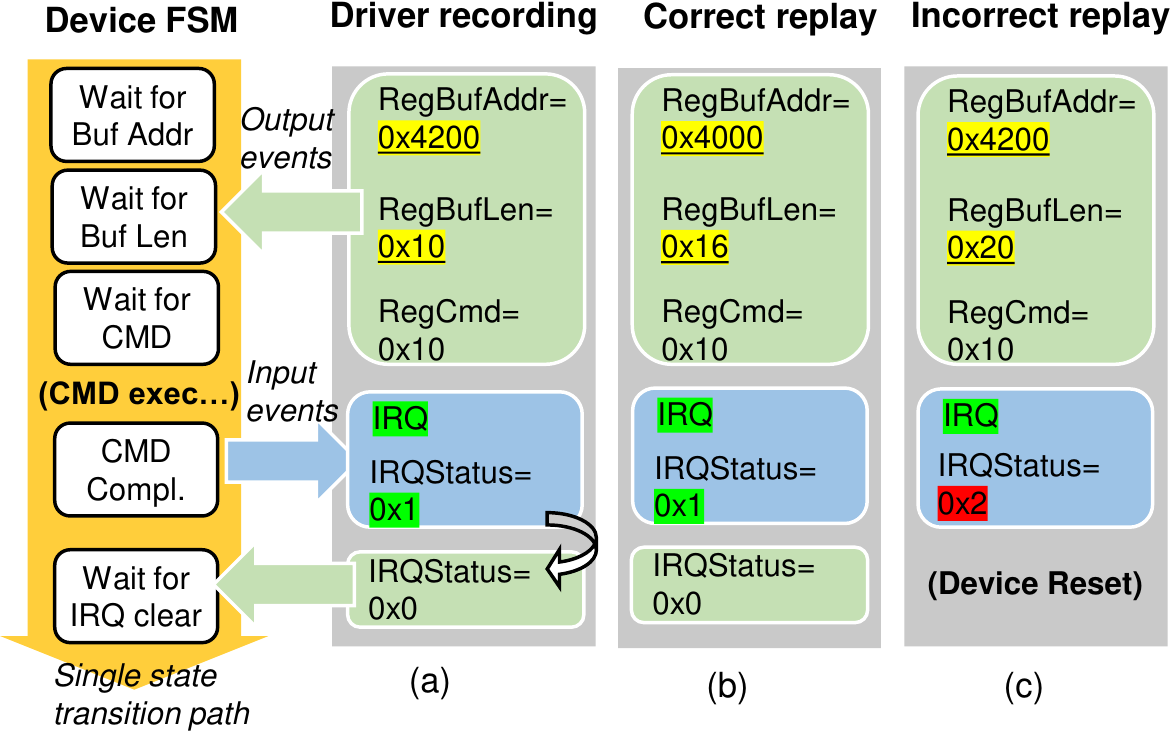}
	\caption{A motivating example of our approach, which captures and reuses a single device state transition path.
	Device FSM is implicitly assumed.
	All input/output events listed are state-changing events. \underline{Underlined values} are generalized in a replay. Highlighted input events in a replay are expected to be matched exactly in a recording. 
	}
	\label{fig:code}
\end{figure}

\begin{myitemize}
	\item 
	\textit{Device FSM.}	
	A device operates its internal finite state machine (FSM), which encodes the driver/device protocol and decides the device behaviors~\cite{rpi3-sdhost,usb-driver,software-fsm}. 
	The FSM is reactive to requests submitted by a driver, e.g. read/write  flash blocks;
	it does not initiate requests autonomously. 
	During request execution, the device state transitions are independent of IO data content, e.g. the state of a MMC controller is unaffected by the block contents being read. 

	We are agnostic to FSM internals: 
	we only assume a device to have an internal FSM, but not the knowledge of the FSM's specifications. 
	This is because devices often have unpublished FSMs with states not exposed to software explicitly.

	\item 
	\textit{Driver/device interactions.}
	The driver's interfaces to a device include registers, shared memory, and interrupts. 
	The interactions are a set of input and output events from the driver's perspective.
	The input events include reading the device registers/shared memory, receiving interrupts from the device, and requesting services from the environment (e.g. DMA memory allocation);
	the output events include writes to the device registers and the shared memory. 

	\item 
	\textit{State-changing events.} They are input/output events used by a driver to shepherd the device FSM executions. 
	Figure~\ref{fig:code}(a) shows a simple example:
	the output events prepare a buffer and kick off the FSM execution of command \code{0x10};
	an input event, i.e. read of interrupt (IRQ) status, reflects the FSM execution outcome;
	seeing the success status (\code{0x1}), the driver de-asserts the IRQ by writing \code{0x0} to the status register with an output event.
	Specifically, we define state-changing events as follows:
	\begin{myenumerate}
		\item 
		All output events.
		They directly change the device state, kick off a hardware job, etc.
			
		\item 
		A subset of input events, including interrupts, service responses from the environment, as well as register/shared-memory reads that have causal dependencies with at least one subsequent output event.
	\end{myenumerate}
	
Note that we define state-changing events broadly, so that we err on the side of falsely assessing a successful replay as a failure, rather than a failed replay as a success. 
The former can be overcome with re-execution while the latter will cause silent errors. 
Therefore, our assessment of replay success is sound. 	
Our system automatically identifies the state-changing events, as will be described in \sect{record}.

\end{myitemize}

\paragraph{The gold driver}
We assume that the full driver implements sufficient state-changing events, so that it can assess if the device has finished the state transitions needed by given requests. 
We do not rule out other bugs in the driver, 
of which consequences will be discussed in \sect{eval}. 






\subsection{Our approach}
\label{sec:bkgnd:approach}

Our idea is  to selectively reuse driver/device interactions induced by device state transitions. 
The core mechanism is as follows. 

%

\begin{myitemize}
\item 
\textit{Design prerequisite}. 
We require that a device always follows the same path of state transitions to finish a given request. 
As such, we configure the driver so that it constrains the device's state space: 
disabling irq coalescing, concurrent jobs, and runtime power management.

\item 
\textit{To record}. 
The driver is invoked with a concrete request, 
e.g. to read 16 blocks starting from block 42. 
In this process, the recorder logs the driver/device interactions, including accesses to registers, the DMA memory, and interrupts. 

\item 
\textit{To replay}.
The recorded log is generalized as an interaction template. 
The template strikes a balance between simplicity and expressiveness. 
It dictates a linear sequence of input/output/meta events, which are the replayer's minimum activities to fulfill the recorded IO jobs. 
The template accepts dynamic inputs (from program/environment/device) which are much broader than just the logged input values. 


\end{myitemize}

\sect{record} and \ref{sec:design:replay} will present the mechanism in full. 

\subsection{Why driverlets work}
\label{sec:design:theory}

We discuss three key questions tied to the \teedriver{} design: 
how does a \teedriver{} assure the correctness? 
how expressive is it?
what happens if it fails?

\paragraph{A \teedriver{} is as correct as a gold driver, as long as the replay is faithful.}
For the replay to be faithful~\cite{r2}, the replayer observes the same sequence of state-changing events as the gold driver observes at record time. 
This can be proven in two steps.
1) The recorded state-changing events constitute a device state transition path;
a faithful replay drives the device through the same path, which is equivalent to the recorded one as far as the device FSM is concerned.
2) If there exists an alternative path undetected by a \teedriver{}, e.g. for a write request the same state-changing events are observed but data is silently lost, the path would also be undetected by the gold driver;
this means the recorded gold driver does not implement sufficient state-changing events for inferring device states, contradicting to our assumption (\S\ref{sec:bkgnd:model}). 

Figure~\ref{fig:code} (b) shows a faithful and correct replay: all the state-changing events match those in the record run (a), except for the parameters in the output events. 

\paragraph{A \teedriver{} is as expressive as a subset of full driver functionalities.}
The subset of functionalities is encoded in device state transition paths, recorded as sequences of state-changing events. 
Through generalizing the output events, a \teedriver{} can vary the stimuli to the device FSM while still staying on the same state transition paths.
It retains the capabilities to access arbitrary IO data but must access the data in ways specified by the recorded paths, 
e.g. accessing at the granularities of 1, 8, or 32 flash blocks. 
It leaves out other driver functionalities, such as accessing data at arbitrary granularities, optimizations for large transfers, and dynamic power management. 



\paragraph{A \teedriver{} deals with state divergence by device reset.}
Should the replayer observe any state-changing event mismatching the recording, 
it deems a divergence in the state transition and a failure in IO jobs, hence an incorrect replay. 
Figure~\ref{fig:code} (c) shows the example where the IRQ status (\code{0x2}) mismatches the recording (\code{0x1}).
Fundamentally,
such divergences happen because the device FSM has received unexpected stimuli. 
We identify three major sources. 
(1) Residual states left from prior IO jobs.
For example, if a device's FIFO is yet to be flushed, 
the replayer may read different numbers of empty FIFO slots at different times.
(2) Fluctuation in chip-level hardware resources such as power, clock, and memory bandwidth. 
(3) Unexpected hardware failures. 
For example, a media accelerator loses the connection to the image sensor.
To prevent divergences (cause 1 above) and recover from transient failures (cause 2 and 3), the \teedriver{} resets the device before executing each template and upon the occurrence of a divergence. 
\sect{eval} will discuss the efficacy of recovery. 
\subsection{Limitations}
\label{sec:bkgnd:limit}


First, 
driverlets' simplicity hinges on the data-independence of device FSM.
As such, driverlets give up on devices where state transitions depend on IO data \textit{content}, a behavior commonly seen on network interface cards (NICs).
An example is BCM4329~\cite{bcm4329}, 
a WiFi NIC that implements 802.11 MAC in its firmware. 
Its Linux driver sends different commands to the NIC based on the header content of received WiFi packets. 
Fortunately, protecting NICs with secure IO is not as crucial as other devices, because security-sensitive network traffic through NIC can be protected with end-to-end encryption. 

Second, 
driverlets' correctness relies on the full driver being gold (\S\ref{sec:bkgnd:model}). 
The assumption is based on our empirical observation on the drivers in the mainline Linux. 
We, however, do not certify the full drivers. 
Doing so would require involving the device and driver vendors. 





%
%
%
%


\section{Record}
\label{sec:record}

\paragraph{How to use}
To generate a \teedriver{}, e.g. for MMC, developers launch a \textit{record campaign}, 
exercising the driver in multiple runs. 
In each run, the developers supply a different sample input (e.g. $blkcnt=1, rw=0$); 
from the run, a recorder produces an \template{}. 
Once done, the recorder signs the templates which are thereafter immutable, and reports a cumulative coverage of the input space, e.g. $0<blkcnt<32, rw=\{0|1\}$. 
Note that the sufficiency of coverage is determined by the developers.
If developers see a desirable input $v$ uncovered, they do a new record run with input $v$ to extend the input coverage. 
They conclude the campaign upon satisfaction of the coverage. 



\subsection{Problem formulation}
\label{sec:record:formulation}

\paragraph{Record entry}
is the entry point of a recording, which requests the driver to complete an IO job. 
A record entry may invoke multiple functions in a driver, 
e.g. a series of \texttt{ioctl()} to acquire an image frame. 


\paragraph{Recorded interfaces} include the following. 

\begin{myitemize}
\item 
\textit{Program $\leftrightarrow$ Driver}: the interface seeds the recording. It includes record entries invoked with concrete arguments, e.g. $\{blkcount=16, blkid=0\}$
 . 


\item 

\textit{Environment $\leftrightarrow$ Driver}: it includes a number of kernel APIs invoked by a driver. 
Examples include DMA memory allocation, random number generation, and timekeeping.


\item 
\textit{Device $\leftrightarrow$ Driver}: the interface is the frontier of driver/device interaction. 
It includes access to device reigsters, descriptors in shared memory, and interrupts.  
\end{myitemize}

\paragraph{Record outcome: \template{}s}
A template exports a callable interface to the replayer. 
The interface has the same signature as a record entry. 
The template comprises a sequence of events in Table~\ref{tab:replay_actions}:
\begin{myitemize}
	\item  
	An input event $V=<I, C, A>$ expects an input $V$ from the interface $I$; $I$ can be the address of a device register, a shared memory pointer, or an environment API (e.g. \code{dma\_alloc}). 
	When the expected value V is specified, it must satisfy the constraint C; 
	otherwise the replayer will reject the input event as a replay failure. 	
	The argument $A$ specifies the input's properties, e.g. expected input length; it can be concrete or a symbolic expression of an earlier input. 
	
	\item 
	An output event $<I,V>$ writes a value V to an interface I. 
	V can be concrete or a symbolic expression of an earlier input. 

	\item 
	A meta event is $delay$ or $poll<I, E, Cond>$. 
	The latter polls from an interface $I$ until a termination condition $Cond$ is met.
	The loop body $E$ is a series of input/output events.	
\end{myitemize}
We select these events as they are generic primitives in the kernel driver framework and are applicable to all driverlets.
For debugging ease, each event is accompanied by its source location in the full driver. 


\begin{table}
	\centering
	\includegraphics[width=0.48\textwidth{}]{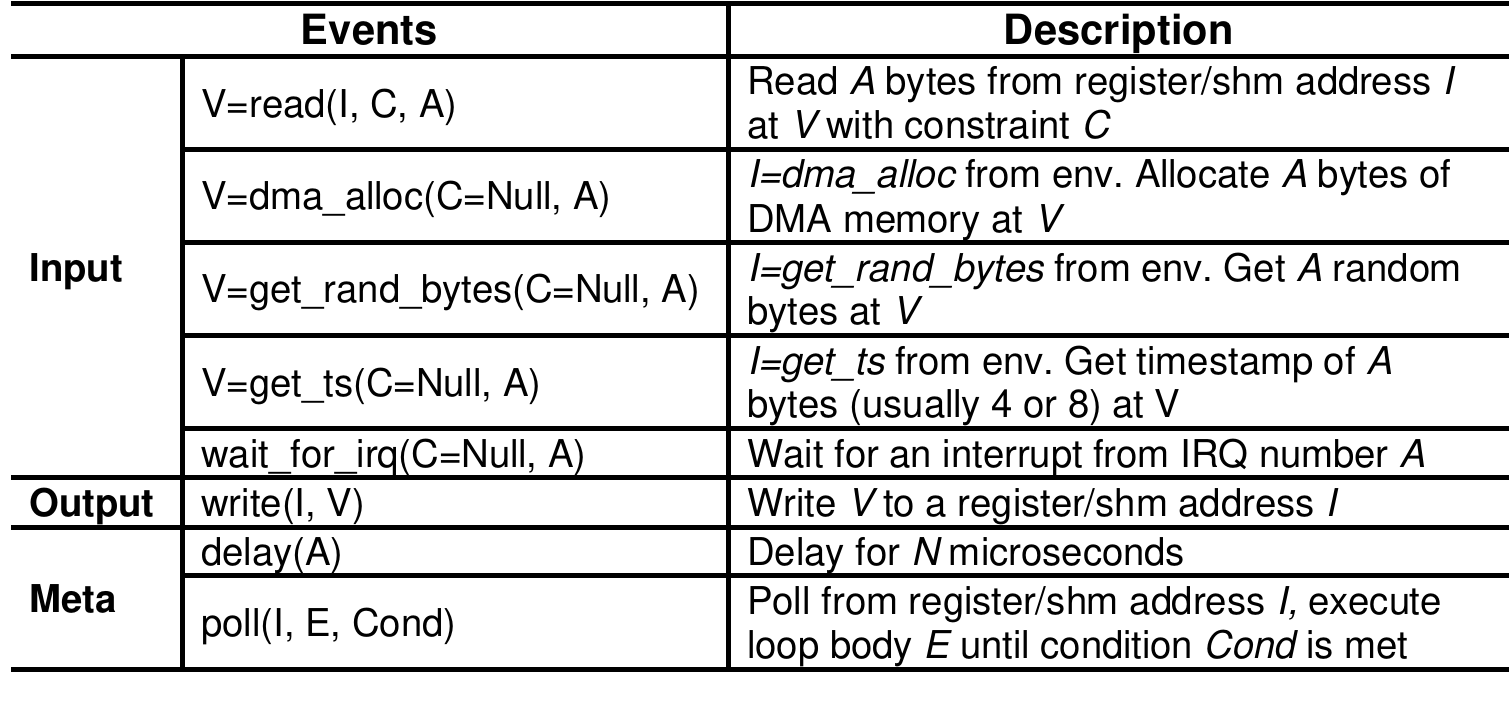}
	\caption{Events in interaction templates for replay.
	They are generic primitives for all driverlets.}
	\label{tab:replay_actions}
\end{table}

\subsection{Key challenges \& solutions}
\label{sec:design:sol}

We next discuss automatic generation of \template{}s, 
which must handle input variation while still maintaining correctness. We have addressed three challenges.


\begin{figure}[h]
	\centering
	\includegraphics[width=0.45\textwidth{}]{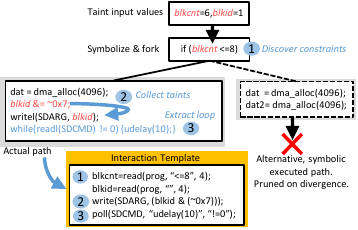}
	\caption{An example of our system extracting constraints, data dependencies, and polling loops into a template.}
	\label{fig:exec-path}
\end{figure}

\paragraph{Challenge I: How to discover causal dependencies between input/output events?}
This is to identify state-changing events (\S\ref{sec:design:theory}) and discover constraints, thus to reject inputs that will compromise replay correctness, i.e. deviation from the correct device state transition path.
Naively matching concrete values from record runs does not work because variations in input values may not indicate state changes.
For instance, input values from a FIFO statistics register are time-dependent and hence do not always correspond to a device state change.
Instead, it's whether subsequent output events causally depend on the FIFO register value that indicates a state change.


\textbf{Solution: Selective symbolic execution} 
Our idea is to study whether variations in input values impact the driver output events, i.e. a causal dependency.
The rationale is the driver, by design, always reacts to state-changing inputs and decides output events correspondingly.
An example is the FIFO statistics register mentioned above:
finding the FIFO watermark too high, the driver writes to a configuration register to tune the bus bandwidth which changes the device state. 




To this end, the recorder uses selective symbolic execution (also called conclic execution~\cite{s2e-hotdep}) to explore the driver's multiple execution paths and assess if they lead to different device state transitions.
To avoid path explosion, the recorder prunes as soon as there is divergence in the output event sequences.
This is shown in Figure~\ref{fig:exec-path}.
As the driver executes with a concrete input $blkcnt =6$ and encounters a conditional branch ($blkcnt \leq 8$), the recorder forks the driver execution, explores both (one actual path with concrete $blkcnt =6$ and an alternative path when $blkcnt > 8$), and compares their subsequent device state transitions.
It discovers that the alternative path has an additional DMA memory allocation, which is a divergent input event as defined in~\S\ref{sec:bkgnd:model}; 
the recorder hence concludes that this path's state transitions are different and prunes it.
Due to such causality, it flags the input event of $blkcnt$ as a state-changing event.
Throughout the execution, the recorder flags state-changing events and collects path conditions, e.g. $blkcnt \leq 8$. 
These path conditions serve as the constraints that input events must satisfy to stay on the same device state transition path.
\sect{record:impl} will present implementation details.

\paragraph{Challenge II: How to discover data dependency between input/output events?}
Input values may be processed and used as an argument of another input event or as an output value by the driver, resulting in data dependencies. 
For instance, upon the notification of an incoming image frame of size $S$, the driver requests a DMA memory region of size $S$, and writes the region's aligned address to a device register.
\textbf{Solution: Dynamic taint tracking}
The recorder discovers data dependencies with dynamic taint tracking: it taints all input values at all interfaces, propagating the taints in driver execution and accumulating both arithmetic and bitwise operations on the taint value until the taints reach their sinks. 
For each taint sink, the recorder replaces the concrete value with a symbol as the taint source with the accumulated operations on the symbol. 
Figure~\ref{fig:exec-path} shows an example.
The recorder tracks $blkid$ and discovers its taint sink $SDARG$ and a bitwise  operation ($blkid\&=\sim0x7$) for alignment.
It hence emits an output event with a $blkid$ symbol plus the operation.

We also face challenges from higher-order dependencies.
As shown in Figure~\ref{fig:exec-path}, the driver allocates more descriptors when $blkcnt>8$. 
Depending on the number of descriptors, a driver, per the device protocol, often links descriptors as pointer-based structures such as lists or arrays of lists; 
it may further optimize the structures based on descriptor addresses, e.g. coalescing adjacent ones. 
\begin{figure}[t]
\centering
	\includegraphics[width=0.48\textwidth{}]{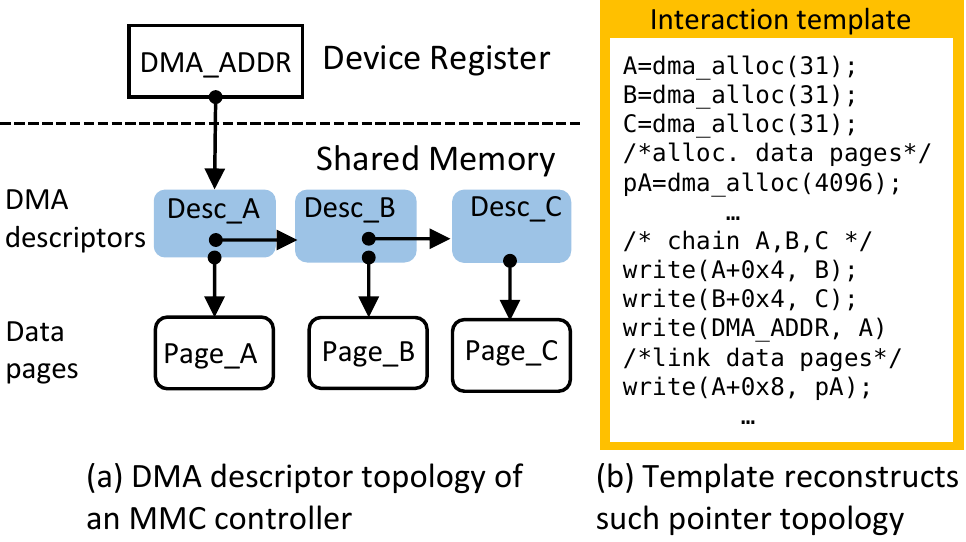}
	\caption{
	To reconstruct a complex descriptor topology (a),	our system mandates a fixed number of DMA allocations in a template (b).
	}
	\label{fig:example-descriptor}
\end{figure}

Figure~\ref{fig:example-descriptor}(a) shows the descriptor topology for an MMC controller (details in Table~\ref{tab:plat}).
For every eight blocks of a request the driver allocates a 4K page and associates a descriptor with the page;
it links them via a physical pointer field in the descriptor and writes the address of the head to DMA\_ADDR register. 

While it may be possible to extract such logic with some device-specific heuristics, such heuristics is likely brittle; 
both the recorder and the replayer will be more complex in order to encode and interpret the logic.
For simplicity, the recorder sets DMA allocation 
as state-changing, mandating that a template must allocate the same number of descriptors as in the record run.
The \template{} in Figure~\ref{fig:example-descriptor} (b) shows a faithful reconstruction of the descriptor topology: the template allocates a fixed number of descriptors, and chains them by writing their symbolized addresses to the corresponding descriptor fields. 
\paragraph{Challenge III: How to record polling loops? }
A major source of nondeterminism is polling loops. 
For example, a driver waits for a command to finish by polling a status register; the number of register reads depends on the timing of command execution.
While conceptually simple, 
a polling loop is known difficult to dynamic code analysis, 
as it can generate many or even infinite alternative executing paths~\cite{loop-problem}.
To explore all paths is impractical. 

\textbf{Solution: Static Loop analysis}
The polling loops in the driver/device interfaces are often succinct, local, and have a clean code structure.
For example, the \rpi{} MMC driver implements polling loops by either using standard register polling functions (e.g. \code{readl\_poll\_timeout}) or a short while loop (<10 SLoC).
With static code analysis~\cite{dr.checker},
we find the polling loops and lift each loop as a standalone meta event, which preserves the loop condition and the input/output events inside the loop body. 
This allows the replayer to execute a varying number of input/output events for a loop.

\section{Replay}
\label{sec:design:replay}

%
%


\paragraph{Overview}
In TEE, a trustlet statically links the replayer and the compressed \template{}s as a library, which constitute ``driverlets'' for target devices. 
To use a \teedriver{}, the trustlet invokes the callable interfaces exposed by the \template{}s (\S\ref{sec:record:formulation}). 
Under the hood, the replayer dynamically selects a template, instantiates it, and executes its input/output/meta events;
the replayer resets devices between template executions and upon any device state divergence.
 

\paragraph{Selecting an \template{}}
The replayer decompresses the \template{} package within TEE.
Upon trustlet invocation, 
the replayer selects one template that has all constraints satisfied by the trustlet inputs.
By design, no two templates can be selected simultaneously; 
otherwise they should have been merged by the recorder on the same state transition path (\S\ref{sec:design:sol}).
If no template is selected, the replayer reports an error that the given inputs are out of coverage. 

\paragraph{Instantiating the template}
The selected \template{} preserves the symbolized input/output values and refers to them by unique names.
Doing so parameterizes the new inputs supplied by the trustlet and allows them to reconstruct the recorded data dependencies.
For physical device addresses, the replayer replaces them with newly mapped TEE virtual addresses.
It updates the callbacks of all events, pointing them to the respective TEE APIs.
Most events do not need special support from the TEE kernel. 
For instance, \code{read} and \code{write}, which constitute over 90\% of all events, are directly dispatched to TEE's memory read/write.
Meta events are implemented as simple loops and delays.
Only input events at the \textit{Env$\leftrightarrow$Driver} interfaces need more environment support, e.g. DMA allocation backed by a CMA pool. 
Luckily, they are often already implemented by existing TEE kernels, which we will describe in \sect{impl}. 

\paragraph{Executing events}
The replayer uses a single-threaded, sequential executor. 
It maintains two contexts: 
a normal context corresponds to the kernel context in the original driver; 
an interrupt context corresponds to the IRQ handler, where only a minimal part is handcrafted to recognize IRQ sources and the rest is for replaying.
The scheduling of two contexts is triggered by the \code{wait\_for\_irq} input event.
The design constrains the device state space in the same way as a record run by limiting hardware concurrency (\S\ref{sec:bkgnd:approach}), hence preventing many potential state divergences. 

The executor is transactional. 
In a successful execution, all input events' constraints must be satisfied, including timely interrupts;
it returns to the trustlet the requested data, if any.
Otherwise, it soft resets the device and re-executes the template, which we next discuss. 




\paragraph{Resetting device states}
The replayer soft resets the device under two circumstances: 1) between \template{} executions, 2) upon device state divergence.
The soft reset brings the device back to a clean-slate state -- as if the device just finishes initialization in the boot up process.
The soft reset recovers from transient device errors. 
In case of persistent divergence despite of soft reset, the executor aborts and dumps the call stack; 
it does so by reporting all previously executed events and their recording sites (source files and line numbers). 
Section~\ref{sec:eval} will evaluate the reset efficacy and its overhead.

\paragraph{Self security hardening}
The replayer hardens itself by implementing a list of stringent security measures:
it verifies recording integrity by developers' signatures; 
it only takes inputs from the trustlet;
it does pervasive boundary checks (e.g. device physical address) on \template{}s and trustlet inputs to mitigate attacks exploiting memory bugs;
it eliminates concurrency to avoid race conditions.
Section~\ref{sec:eval:analysis} will present a security analysis. 

\section{Implementation}
\label{sec:impl}

\subsection{Recorder}
\label{sec:record:impl}

We implement the recorder in 2K SLoC C code based on S2E~\cite{s2e}, a popular symbolic execution engine. 
We choose it because it provides in-house support for analyzing Linux kernel drivers and is based on QEMU~\cite{qemu}, whose dynamic binary translation (DBT) engine enables us to trace driver execution at the instruction granularity. 
We use existing \code{LoopDetector} S2E plugin to extract the polling loops.
We next focus on the recorder implementation details of selective symbolic execution and dynamic taint analysis, which primarily relies on DBT.
We omit the details of DBT lingoes; 
an interested reader may refer to~\cite{qemu}.
\paragraph{Matching state transition paths}
As described in \sect{design:sol}, we first symbolize all input events, including arguments from record entries, and register/shared-memory addresses (e.g. 24 addresses for MMC, which we will describe shortly in \sect{case}). 
We do so by first annotating their corresponding kernel sources, each with a custom CPU instruction.
When the DBT engine translates the custom CPU instruction, it traps the execution and examines any path condition on the symbol;
if any, the recorder logs the path condition and forks two new translation blocks, one as the current execution path with the concrete value and the other as the alternative symbolized execution path;
\sect{case} will present the logged path conditions.
The recorder maps the state transition paths into the sequence/chain of translation blocks.
As long as record runs follow the same sequence of translation blocks, the recorder deems them undergo the same state-changing events and hence are on the same state transition path. 
To ensure a complete and correct driver execution of the DBT engine, we supply it with concrete input values of device registers, which are collected from the  of a \rpi{} running the same record campaign side-by-side. 
A similar practice is done by Charm~\cite{charm}.
To save time and avoid path explosion, the recorder emulates kernel API invocations which are input events at \textit{Env$\leftrightarrow$Driver} interface, e.g. \code{dma\_alloc}.
The recorder does so by checking the function addresses being translated; 
upon meeting them, the recorder logs their arguments (e.g. DMA allocation size) and returns directly with a symbolized result (e.g. DMA address, timestamp), instead of translating the actual kernel functions as-is.

\paragraph{Collecting input event taints}
To implement dynamic taint analysis, we interpose on the instruction translation, similar to~\cite{decaf++}.
The recorder inserts Tiny Code Generator (TCG) IR for each translated instruction which checks the taint status of source operand.
The recorder applies taint propagation rules similar to~\cite{taint-rule}; 
depending on the rule, it updates the taint status for the destination operand.
Meanwhile, the recorder logs the taint operations as its corresponding C code for debugging ease. 
In practice, we have found an input event is usually tainted for only a few times before reaching its sink as an output event, which is often a device register;
\sect{case} will present them.

\subsection{Replayer}
We implement the replayer in 1K SLoC within OPTEE-OS~\cite{optee}, whose key responsibility is to execute the replay events in an interaction template with new, dynamic values at runtime. 
For simplicity, we emit the recorded interaction templates as standalone header files, consisting of human-readable input/output/meta events for debugging ease;
each event is encoded as a function, whose signature is listed in Table~\ref{tab:replay_actions}. 
For instance, a read event from SDCMD register expecting a 0x0 value manifests as \code{read(SDCMD, "=0x0", 4)}.
We hence statically compile the templates and links them with the replayer.
The replayer implements the events as follows.
It implements read/write as uncached memory access to ensure device memory coherence;
it implements poll/delay events as while loops, which continuously check against the termination conditions or timeout.
For the rest events that are more complex, the replayer leverages the existing OPTEE-OS facilities: for DMA allocation, it uses the default OPTEE-OS memory allocator (i.e. \code{malloc}), which already allocates contiguous pages;
OPTEE-OS also implements hardware RNG to for random bytes and RPC to normal world for getting timestamps.


\section{Experiences}
\label{sec:case}

We put ourselves in the shoes of developers and apply our system to a variety of devices: MMC, USB, and the CSI camera.
We select them because they have important use cases of secure IO and their drivers are complex and known difficult for TEE.
For each driverlet, we as developers write only 20 SLoC in C as a record campaign to exercise the driver execution;
we also record the device initialization process in the driver loading phase.



\begin{table}[h]
\centering
	\includegraphics[width=0.47\textwidth{}]{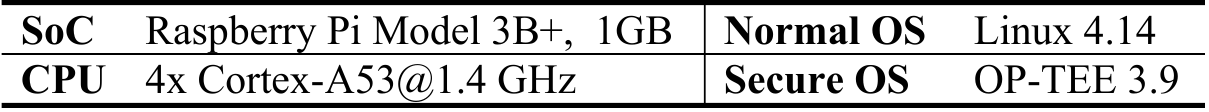}
	\includegraphics[width=0.48\textwidth{}]{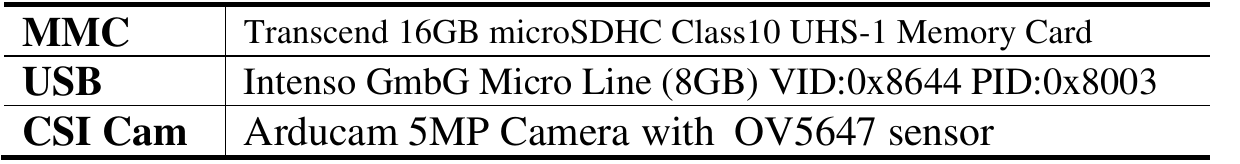}
	\caption{The test platform and peripherals used.}
	\label{tab:plat}
\end{table}

Our test platform is RPi3;
Table~\ref{tab:plat} shows the details of the board and peripherals.
We choose RPi3 because of its popularity, good support by open-source TEE and the QEMU emulation.
As we will show, despite RPi3's high popularity and an active developer community, building TEE drivers with existing approaches is nevertheless challenging. 


For each driver, we report our findings and answer the following questions: 
\begin{myitemize}
\item How complex is the driver and why is it complex? 

\item 
With hindsight,  what are discovered by our toolkit and what interactions are recorded?
\end{myitemize}

\subsection{MMC}
\label{sec:case:mmc}

\subsubsection{Driver overview}
MMC is a common interface to off-chip flash, e.g. SD cards and eMMC. 
The Linux MMC framework supports more than 20 MMC controller models with diverse interfaces and poor documents. 
The framework abstracts the common driver logic as a ``core'' with 15K SLoC in 40 files; 
the driver for a concrete MMC controller plugs in the MMC core via a wide interface, including 44 callbacks implemented in four structs.
The driver itself often has a few K SLoC. 
All combined, the MMC framework implements an FSM with thirteen states and hundreds of transitions among them.
The FSM supports rich features, including streaming access of blocks and medium hotplug. 

\subsubsection{Recording outcome}
\begin{table}[h]
	\includegraphics[width=0.47\textwidth{}]{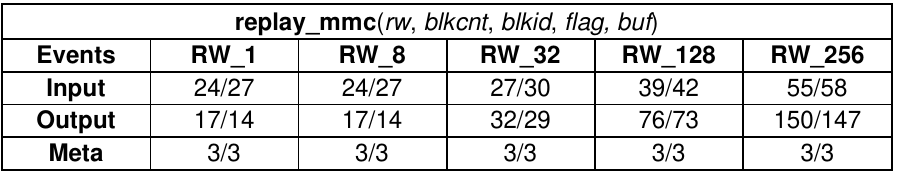}
	\caption{Breakdown of 10 interaction templates of MMC given the replay entry \textit{replay\_mmc}. 
	RD/WR templates of same $blkcnt$ merged in one column (e.g. RW\_1), separated by ``/'' (e.g. 24, 27 input events for RD\_1, WR\_1 respectively).
	}
	\label{tab:mmc:replay_action}
\end{table}

%
%

We choose to implement a record campaign of 10 requests: read/write of 1, 8, 32, 128, 256 blocks.
We reserve the 15-th DMA channel for recording.
The replay entry, 10 automatically generated templates and their events breakdown are listed in Table~\ref{tab:mmc:replay_action}.
Each template covers the full range of 31M blocks (512 bytes each) available.
We have found templates are similar with each other, e.g. RW\_8 and RW\_32 only differ by 2 DMA allocations, due to additional descriptors. 

\subsubsection{Post analysis}
Our system has observed two distinct state transition paths w.r.t different flags. 
(1) if $O\_DIRECT$ is specified, the full driver shifts individual words of data blocks from/to the SDDATA register. 
(2) otherwise the driver uses DMA transactions to move the data. 
Notably, even when using the DMA transaction, the driver moves the last 3 words via SDDATA on read path. 
This seems to work around an undocumented bug in the SoC's DMA engine, which cannot move the last few words of a transfer. 

Each \template{} involves 15 different registers out of 24 total registers of MMC controller and a system-wide DMA engine, in three groups: 8 for configuring the controllers, 5 for sending MMC commands, and 2 for controlling the DMA engine. 
Key symbolized input values are encoded into 32-bit words written to 4 different registers, as shown in Table~\ref{tab:mmc:record}.
Notably, 
including read/write, five different SD commands (CMD17, 18, 23, 24, 25) are sent to the SDCMD register; 
CMD23 (set block count) is used on the read path but not write path.
Our system has also gathered taints on \textit{blkid} by Linux block layer for 8-block alignment.
We tried manually feeding misaligned \textit{blkid}, which has caused state transition divergence.

\begin{table}[h]
	\centering
		\includegraphics[width=0.48\textwidth{}]{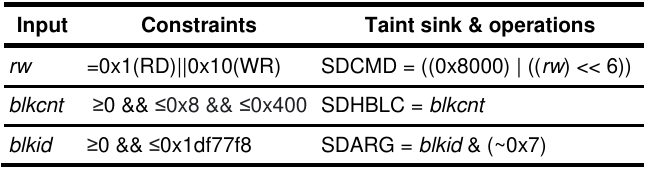}
		\caption{Key constraints and taint operations of inputs on the RW\_1 template of the MMC driverlet. 
		}
		\label{tab:mmc:record}

\end{table}
\subsection{USB}
\label{sec:case:usb}

\subsubsection{Device overview}
USB serves as a common transport between CPU and diverse peripherals, e.g. keyboards, and flash drives. 
A typical USB controller exposes more than 100 registers. 
A device driver programs them to initiate \textit{transactions} and dynamically schedule them on multiple transmission channels;
the controller translates each transaction into up to 12 types of packets on the bus. 
Through 21K SLoC in 82 files, Linux kernel fully implements the driver FSMs, 
which are big in order to accommodate rich features (e.g. dynamic discovery of bus topology) and various runtime conditions (e.g. device speed mismatch, bus checksum errors). 

We focus on USB mass storage for its significance to TEE's secure storage. 
The driver accepts block requests, translates the requests to various SCSI commands, and further maps the commands to USB bulk transactions. 


\subsubsection{Recording outcome}
%
We apply the same record campaign as MMC for 10 read/write requests. 
We reserve the 1st transmission channel.
We disable Start-of-Frame (SoF), which is used to schedule USB transactions proactively for isochronous USB devices and is unfit for our model (\S\ref{sec:bkgnd:model}). 

For the record campaign, our system emits 200-1500 events for 10 \template{}s, each covering the whole 15M blocks of the USB storage.
Interestingly, the number of events are identical in a read template and the corresponding write template;
it appears only some output values to certain descriptors differ. 
\subsubsection{Post analysis}
Our system has captured non-trivial interactions. 
To write blocks smaller than one LBA (4KB), the driver reads back the entire LBA, updates in memory, and writes back. 
The driver also selects the best SCSI commands: 
there exist five variants of a SCSI read/write command, which have different lengths and can encode different ranges of LBA; 
the driver picks the 2nd shortest ones (i.e. read 10, write 10) just long enough to encode the requested LBA addresses. 

Our system identifies 14 USB controller registers out of the 64 KB register range in three categories:
5 manage USB peripheral states (e.g. device power state); 
3 manage the controller itself (e.g. interrupts); 
6 manage transmission channels (e.g. DMA address \& size). 
Unlike MMC, the USB driver communicates with the device primarily via two descriptors: one command block wrapper (CBW) for SCSI commands and the other for querying command status (CSW).  
The data dependencies are similar to MMC, except aligned \textit{blkid} and \textit{blkcnt} are written to CBW instead of registers.
Our system identifies two statistic inputs which are unseen in MMC: a monotonic command serial number and an HFNUM register read.
As they are not state-changing, our system does not impose any constraints. 

\subsection{Camera}

\subsubsection{Device overview} 
On a modern SoC, CPU typically offloads video/audio processing to accelerators, which communicate with CPU primarily via messages backed by shared memory.
We studied VC4, the multimedia accelerator of \rpi{}. 
According to limited information, VC4 implements key multimedia services, including camera input, display, audio output. 
It communicates with CPU via a complex, proprietary message queue called VCHIQ.
Using VCHIQ as a transport, each media service further defines its message format and protocol, e.g. MMAL (MultiMedia Abstraction Layer) for cameras. 
The details of VCHIQ, as well as internals of VC4, remain largely undocumented.

We focus on one media service essential to secure IO: image capture from a CSI camera (a pervasive image sensor interface used in modern mobile devices).

\begin{table}[h]
	\centering
	\includegraphics[width=0.48\textwidth{}]{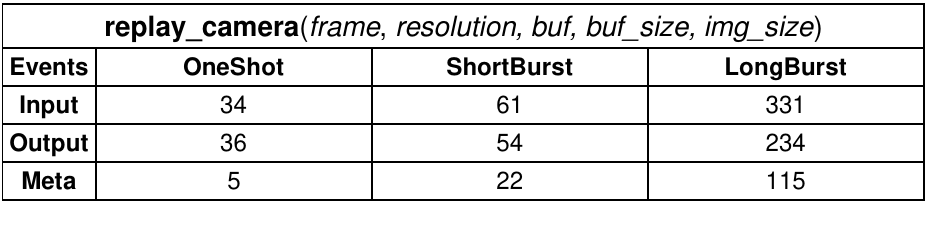}
	\caption{Events breakdown of 3 \template{}s under a given resolution for the CSI camera with the replay entry \textit{replay\_camera}.}
	\label{tab:vchiq:template}
\end{table}

\subsubsection{Recording outcome}
 We choose the record campaign: capture 1, 10, 100 image(s) at 720p, 1080p, 1440p. 

For the record campaign, as listed in Table~\ref{tab:vchiq:template}, our system emits 3 \template{}s (OneShot, ShortBurst, LongBurst for capturing 1, 10, 100 images(s) respectively) of 75-680 events.
Templates cover all resolutions supported by the camera, and a practical range of frames.  
Unlike both MMC and USB, the template's input/output events are mostly accessing shared memories.  

\begin{table}[h]
\centering
	\includegraphics[width=0.48\textwidth{}]{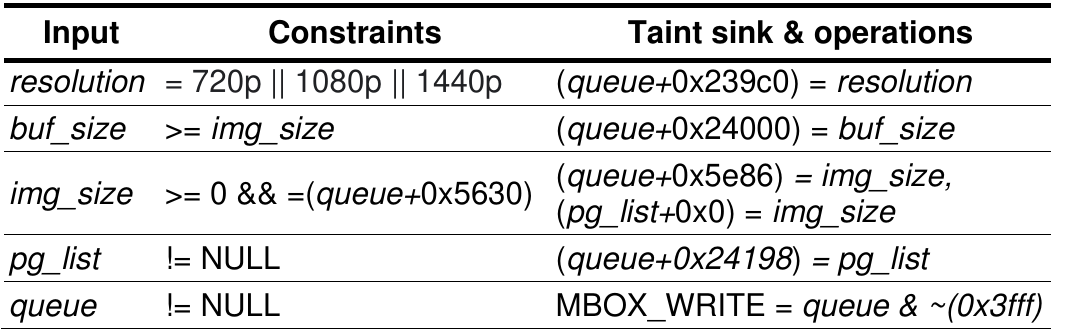}
	\caption{Key constraints and operations of input values for the camera driverlet. 
	\textit{queue} and \textit{pg\_lsit} are DMA addresses allocated from \code{dma\_alloc}.
	}
	\label{tab:vchiq:record}
\end{table}

\subsubsection{Post analysis}
Of the templates, our system identifies only three registers in use.
Two of them are a pair of ``doorbells'' for inter-processor interrupts between CPU and VC4;
only one MBOX\_WRITE register acts as a sink for \textit{queue}, which points to base address of message queue. 
We summarize the discovered dependencies in Table~\ref{tab:vchiq:record}.
A notable constraint is for \textit{img\_size} input value.
It is assigned by VC4 and is sent back to VC4 in a message initiating bulk receive procedure;
later when the procedure finishes, VC4 passes another input value indicating successful transmission size at \textit{queue+0x5630}, which \textit{img\_size} must exactly match. 


Catering to concurrent media services, the message queue has a sophisticated structure. 
It is divided into many 4KB slots, each assigned either to CPU or VC4 for enqueueing messages independently.
Each slot holds multiple messages, ranging from 28 bytes to 306 bytes.
These messages belong to tens of types, either for configuring the device, e.g. opening a service ``port'' of VC4, setting frame resolution; or for moving data, e.g. bulk receive. 
Slot 0 is special, as it contains metadata that describes the whole message queue and will be updated by both CPU and VC4, e.g. the number of slots, slot allocations, read/write locations in the message queue. 
The doorbell registers -- 
BELL0 and BELL2 signal CPU and VC4 to parse new message, respectively. 
Upon a new doorbell, a slot handler thread actively polls and parses the message; a sync thread synchronizes CPU-VC4 shared states in slot 0; a recycle thread actively frees and recycles used slots.

\section{Evaluation}
\label{sec:eval}

In this section we answer the following questions:
\begin{myenumerate}
	\item 
	How does our system reduce developer efforts? (\S\ref{sec:eval:efforts})
	\item 
	Why are \teedrivers{} correct and secure? (\S\ref{sec:eval:analysis})
	\item 
	What is the overhead of \teedrivers{}? (\S\ref{sec:eval:overhead})	
	\item 
	How to use \teedrivers{} to build trustlets? (\S\ref{sec:eval:case}) 
\end{myenumerate}

%
%
%

\subsection{Analysis of developer efforts}
\label{sec:eval:efforts}
%

We compare three approaches to implementing the \textit{same IO functionalities} as described in \sect{case}. 


\begin{table}[h]
\centering
	\includegraphics[width=0.48\textwidth{}]{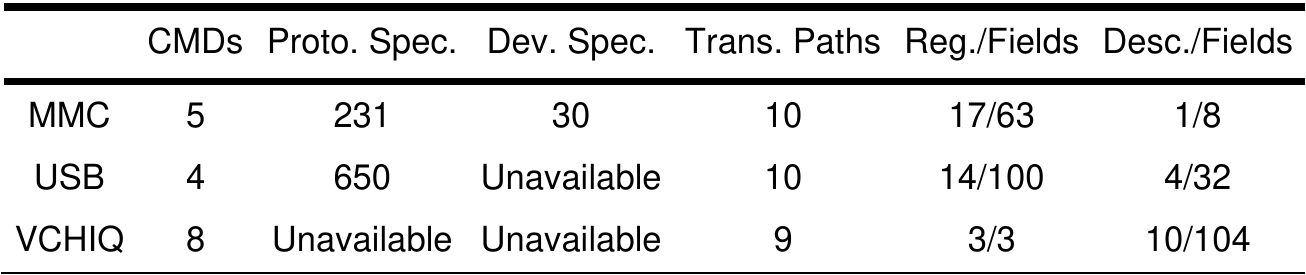}
	\caption{Efforts for building drivers from scratch, showing the needed device knowledge. 
	Proto. Spec. is Protocol Specifications and Dev. Spec. is Device Specifications; both are counted in \textit{number of pages}.}
	\label{tab:eval:build-from-scratch}
\end{table}

\paragraph{\textit{Build From Scratch}}
Developers handpick a set of device commands to implement, 
for which they consult device specifications, implement the state transitions, and work around hardware quirks (\S\ref{sec:case:mmc}).
Table~\ref{tab:eval:build-from-scratch} gives a summary of the needed knowledge.
We estimate that each driver takes a few months to build.  

\begin{table}[h]
\centering
 	\includegraphics[width=0.48\textwidth{}]{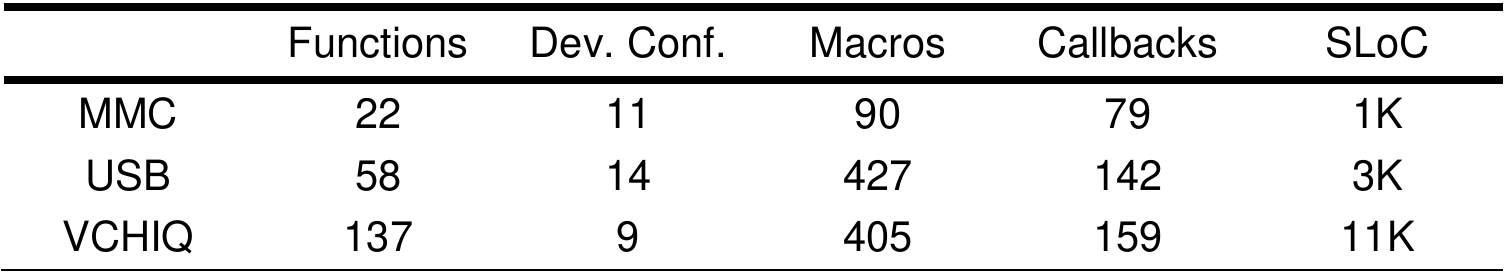}
	\caption{Efforts for porting Linux drivers, showing the code the developers need to reason about and potentially modify.
	Dev. Conf. is Device Configurations.}
	\label{tab:eval:trim-down}
\end{table}

\paragraph{\textit{Port}}
Developers familiarize themselves with device specifications, and decide what driver/kernel functions to port (and what not to).
They must spin off the code paths, which span 22--137 driver functions as we measure. 
To resolve the kernel dependencies of the select code paths, the developers have to port at least 1K, 3K, 11K SLoC for each of the MMC, USB, and VCHIQ drivers; 
they need to port at least 5K SLoC for emulated kernel frameworks such as block, power and memory management.
We estimate that it takes a few months to understand each driver and port its dependencies, plus several months to build the emulated kernel frameworks. 

\paragraph{Our approach}
By comparison, we require much lower developer efforts. 
We build the toolkit in several weeks, which is a one-time effort.
To derive each \teedriver{}, we familiarize ourselves with the full driver's register definitions and instrument the input interfaces for code analysis. 
Each \teedriver{} takes 1--3 days. 
\subsection{Correctness \& Security analysis}
\label{sec:eval:analysis}
We experimentally validate the correctness and security concerns.
We discuss them separately: 
correctness violation is caused by software semantics bugs; 
security breaches are caused by active attackers who compromise the software. 

\subsubsection{Correctness}
Driverlets' correctness can be affected by semantics bugs in the OS and the driver for recording, e.g. the driver writing to a wrong device register. 
Such bugs result in malformed recordings and incorrect replay outcomes. 
Driverlets neither mitigate nor exacerbate such semantic bugs. 
Our recorder and replayer may introduce semantic bugs, e.g. due to implementation glitches.
Yet, we expect such bugs to be rare because of software simplicity: the recorder and the replayer are only 3K SLoC. 

\paragraph{Experimental validation}
We further validate driverlets' correctness experimentally by following the practice in prior work~\cite{symdrive,revnic}.
We develop test scripts to do the following:

\begin{myitemize}
	\item
	\textit{Statically vetting of templates.}
	Our scripts scan templates as a sanity check for the integrity of state-changing events, e.g. which SCSI command is written to what register, what MMAL message is sent. 
	The scripts verify that the templates conform to the record campaign and developer requests. 
	
	\item 
	\textit{Validation of IO data integrity.}
	For MMC \& USB, our scripts verify that values read by \teedrivers{} match those by native drivers and that writes reach the storage; 
	for VCHIQ, the scripts analyze the captured images and verify that they are in the valid JPEG format.
	
	\item 
	\textit{Stress testing templates.} 
	Our scripts enumerate templates to stress test and validate the coverage of input space. 
	The scripts verify a 100\% coverage for MMC/USB blocks (MMC: >31M, USB: >15M of blocks); 
	for VCHIQ, the scripts repetitively invokes templates for 10K times and verify runs produce integral frames.
	
	\item 
	\textit{Fault injection.} 
	We validate that \teedrivers{} handle state divergences properly.
	To do so, we unplug the MMC/USB storage medium amid a replay run for a large data transfer (2K blocks). 
	The \teedriver{} correctly detects divergence and attempts re-execution with reset.
	Because the injected failure is non-recoverable through soft reset, the \teedriver{} eventually gives up. It reports unexpected values from two status registers (SDEDM for MMC and GINSTS for USB) as well as the source lines of the register reads in the original drivers, allowing quick pinpointing of the failure causes. 
	
\end{myitemize}

\subsubsection{Security}
\paragraph{Threat model}
We follow the common threat model of TrustZone~\cite{streamBox-TZ,trustshadow}.
On the target machine: we trust the SoC hardware including any firmware; 
we trust the TEE software; we do not trust the OS. 

We assume that the OS and driver on the developer's machine for recording are uncompromised. 
The rationale is that the developer machines are often part of a software supplychain with strong security measures. 
Compromising them requires high capability and long infiltration campaigns~\cite{supplychain-whitepaper}. 

\paragraph{Security benefits}
By leaving drivers out of TEE, \teedrivers{} therefore keep the TEE immune to extensive vulnerabilities in the driver code. 
Examples include dirtyCoW~\cite{CVE-2016-5195} caused by race conditions in the page allocator, BadUSB~\cite{badusb} caused by unrestricted privileges in the USB stack, and memory bugs in drivers~\cite{CVE-2016-7389, CVE-2016-6775, CVE-2016-9120}. 
They could have been exploited by adversarial peripherals (a malicious USB dongle~\cite{badusb}) or malformed requests sent from the OS to the TEE. 


\paragraph{Attacks against \teedrivers{}}
(1) Fabricating interaction templates is unlikely. 
This is because they are signed by developers, whose recording environment is trusted.
(2) Attacks against the replayer. 
Vulnerabilities in the replayer may be exploited by entities external to the TEE, e.g. an adversarial OS or peripherals. 
Successful attacks may compromise the replayer or even the whole TEE. 
However, such vulnerabilities are unlikely due to replayer's low codebase (only 1K SLoC), simple logic (such as minimalist memory management and well-defined event semantics), and stringent security measures (\S\ref{sec:design:replay}).
\vspace{-2mm}
\subsection{Overhead}
\label{sec:eval:overhead}
\subsubsection{Methodology}

The details of our test hardware are listed in Table~\ref{tab:plat}.
Because the RPi3 board does not implement TZASC, 
we modified \arm{} trusted firmware to assign devices instances to TEE. 
We isolate the whole MMC and VC4 instance.
We reserve 3 MB of TEE RAM and use the stock OPTEE allocator for DMA memory. 


\begin{table}[h]
	\centering
	\includegraphics[width=0.48\textwidth{}]{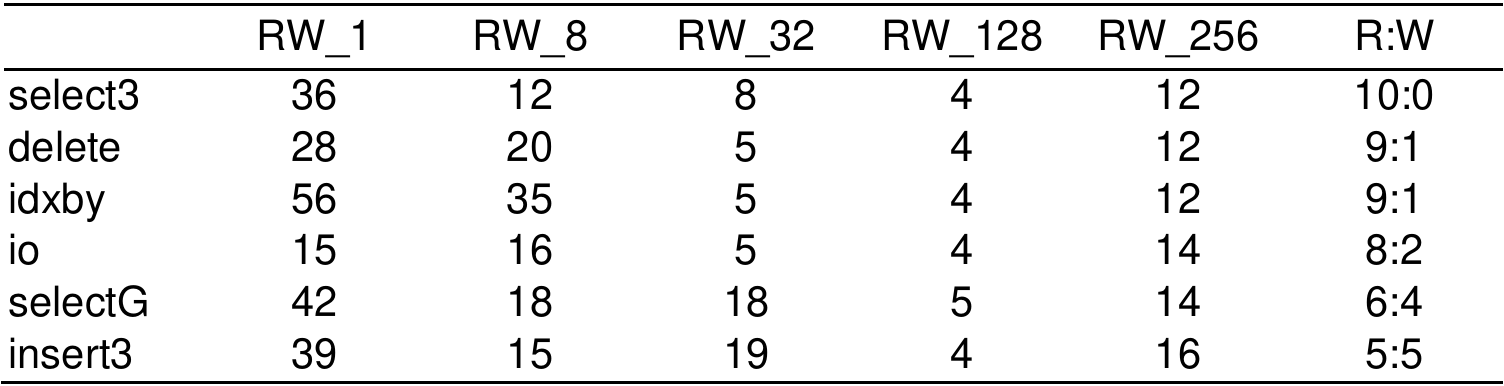}
	\caption{Benchmarks used from SQLite test suites and a breakdown of \template{} invocations.
	Template details are shown in Table~\ref{tab:mmc:replay_action} and in \sect{case}.
	}
	\label{tab:eval:benchmarks}
\end{table}

\paragraph{Benchmarks}
1) \textbf{SQLite-MMC}: we choose SQLite, a popular lightweight database to test MMC.
We pick 6 tests from SQLite test suites to diversify read/write ratios;
breakdown of their template invocations and read/write ratios are shown in Table~\ref{tab:eval:benchmarks}. 
The tests issue their disk accesses in TEE and we report IOPS.
2) \textbf{SQLite-USB}: 
we test USB mass storage with the same SQLite test scripts.
3) \textbf{Camera} (OneShot/ShortBurst/LongBurst): we request VCHIQ to capture 1, 10, 100 still images frames at 720P, 1080P, and 1440P.
We report the latency of each request.




Note that unlike many TrustZone systems~\cite{sbd,trustshadow}, \teedrivers{} do not incur world-switch overheads because they fully run inside the \TZ{}. 

\paragraph{Comparisons}
We compare \teedrivers{} with drivers on Linux 4.14 which finishes the same benchmarks as follows:
for \textbf{SQLite}, we run a test harness invoking the full drivers with the same block accesses with default flags (\textbf{native}) and with an additional O\_SYNC flag (\textbf{native-sync});
for \textbf{Camera}, we run \code{v4l2-ctl} to request the same number of frames at corresponding resolutions (\textbf{native}).

\subsubsection{Macrobenchmarks}

\begin{figure}
	\centering
	\begin{minipage}[b]{0.23\textwidth}
		\centering
		\subcaptionbox{SQLite-MMC\label{fig:eval:mmc-sqlite}}{
		\includegraphics[width=\textwidth]{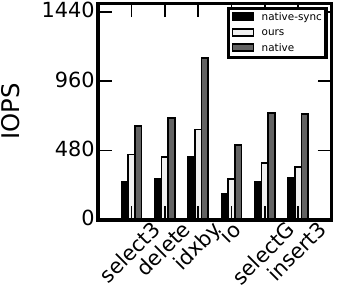}
		}
	\end{minipage}
		\begin{minipage}[b]{0.23\textwidth}
		\centering
		\subcaptionbox{SQLite-USB\label{fig:eval:usb-sqlite}} {
		\includegraphics[width=\textwidth]{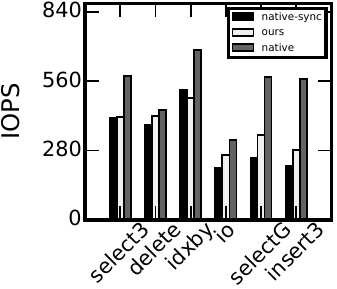}
	    }
	\end{minipage}
	\caption{SQLite benchmarks for MMC and USB \teedrivers{}.
	Driverlets' overhead increases with write ratios due to they mandate synchronous IO jobs while native drivers do not.}
\end{figure}

\paragraph{SQLite-MMC}
Figure~\ref{fig:eval:mmc-sqlite} shows the results. 
MMC \teedriver{} achieves a decent performance: 
on average, 
it achieves 434 IOPS, executing over 100 queries per second. 
As a reference, 
the throughput is a few orders of magnitude higher than secure storage hardware, e.g. RPMB~\cite{rpmb-perf}.

Compared with the native driver, MMC \teedriver{}'s throughput is 1.8$\times$ lower on average. 
The overhead grows with the write ratio, e.g. select3 (read-most) incurs  1.4$\times$ overhead while insert3 (write-most) incurs 2$\times$. 
This is because the \teedriver{} mandates synchronous IO jobs: 
while the native driver does not wait for writes to complete, the replayer must wait to match state changing events.
To validate, we mandate O\_SYNC flag in the native driver execution (native-sync) and measure throughput 1.5$\times$ lower than \teedrivers{}. 
This is because \teedriver{}s forgo complex kernel layers such as filesystems and driver frameworks. 
\paragraph{SQLite-USB}
Figure~\ref{fig:eval:usb-sqlite} shows the results.
The \teedriver{} achieves 369 IOPS which is over 90 queries per second. 
The overhead compared with the native driver is 1.5$\times$.
Such overhead is also caused by synchronous writes, where the write-most workload (insert3) incurs the highest overhead of 2$\times$. 
Native-sync is 1.7$\times$ lower than the native USB driver and 1.2$\times$ lower than USB \teedriver{}.


\begin{figure}[t]
\centering
	\includegraphics[width=0.48\textwidth{}]{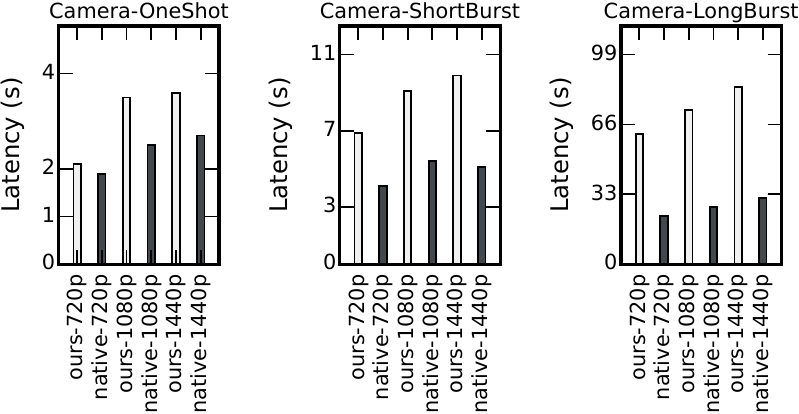}
	\caption{Image capturing latency for Camera benchmarks.
	OneShot, ShortBurst, and LongBurst are for capturing 1, 10, and 100 images(s) respectively.}
	\label{fig:eval:cam-lat}
\end{figure}

\paragraph{Camera}
Figure~\ref{fig:eval:cam-lat} shows the results.
The per-frame latencies of \teedriver{} range from 2.1s (720p) to 3.6s (1440p) which are usable to many surveillance applications that periodically sample images~\cite{diva}.
The per-frame latencies decrease with the number of frames per burst, because for each burst the \teedriver{} pays a fixed cost to initialize the camera and the media accelerator. 
It replays 41 events for the initialization and 5 events to capture each subsequent frame. 

Compared with the native driver, our latency is only 11\% higher for a one-frame burst and is 2.7$\times$ higher for a 100-frame burst.
This is again because the \teedriver{} must wait for individual IRQs as mandated by the templates (\S\ref{sec:design:theory}), while the native driver processes coalesced IRQs. 
As requests contain more events (e.g. 75 vs. 680 to capture 1 and 100 frames, Table~\ref{tab:vchiq:template}), the delays of waiting for IRQs are more pronounced. 

\subsubsection{Microbenchmarks}
\begin{figure}[t]
	\begin{minipage}[t]{0.23\textwidth}
		\centering
		\includegraphics[width=\textwidth]{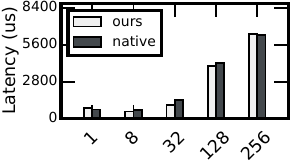}
		\vspace{-2em}
		\subcaption{MMC read latency}
	\end{minipage}
	\begin{minipage}[t]{0.23\textwidth}
	\centering
	\includegraphics[width=\textwidth]{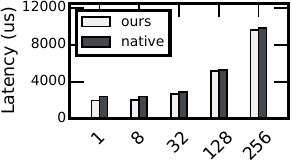}
	\vspace{-2em}
	\subcaption{MMC write latency}
	\end{minipage}
	\begin{minipage}[t]{0.23\textwidth}
		\centering
		\includegraphics[width=\textwidth]{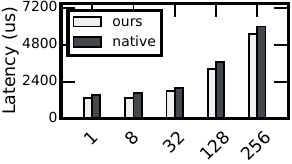}
		\vspace{-2em}
		\subcaption{USB read latency}
	\end{minipage}
	\begin{minipage}[t]{0.23\textwidth}
		\centering
		\includegraphics[width=\textwidth]{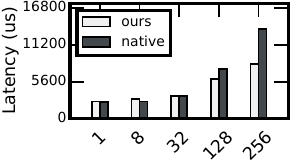}
		\vspace{-2em}
		\subcaption{USB write latency}
	\end{minipage}
	\caption{Microbenchmarks of read/write on the MMC and USB driverlets.
	X-axis: number of blocks, Y-axis: Latency in milliseconds.
	Driverlets achieve near-native performance or even outperform the native 256-block writes due to its simplicity.}
	\label{fig:eval:micro}
\end{figure}

We show latencies to execute individual templates for MMC and USB in Figure~\ref{fig:eval:micro}. 

In both reads/writes, the \teedriver{}'s latencies are near-native or even slightly lower than the native ones (12\% and 13\% lower for MMC and USB respectively).
On larger block writes (e.g. 256), the latency of USB \teedriver{} is even 40\% lower;
this is due to the \teedriver{}, unlike a native driver, does not run transfer scheduling logic for individual 4KB data pages. 
This confirms observations in the macrobenchmarks, where the \teedriver{} outperforms native-sync. 


\subsubsection{Memory overhead.}
The \teedriver{} executables for MMC, USB, and VCHIQ are of 
6 KB, 26 KB, and 19 KB, respectively. 
For implementation ease, our current recorder emits templates as human-readable documents. 
Conversion to binary forms is likely to further reduce their sizes. 

\subsection{End-to-end use case}
\label{sec:eval:case}
\begin{figure}[t]
\centering
	\includegraphics[width=0.48\textwidth{}]{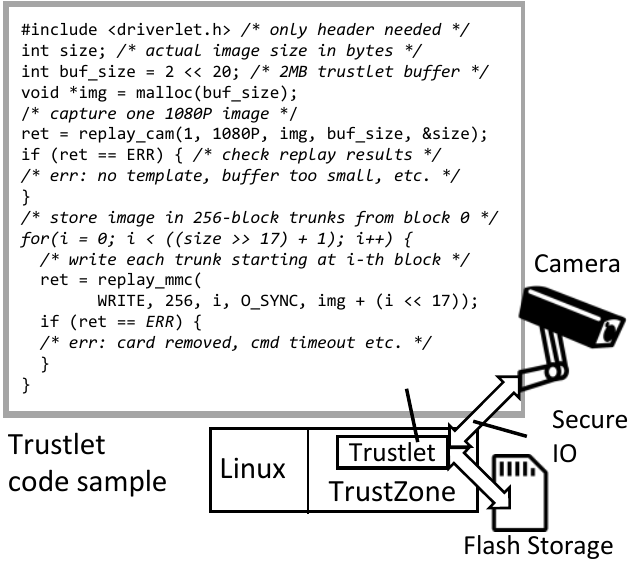}
	\caption{A trusted perception trustlet built atop \teedriver{}s, which expose two simple interfaces (\textit{replay\_cam} and \textit{replay\_mmc}).
	}
	\label{fig:eval:use-case}
\end{figure}

To showcase the use of \teedriver{}s, 
we build an end-to-end trustlet for secure surveillance in only 50 SLoC and less than an hour.
As shown in Figure~\ref{fig:eval:use-case}, 
the trustlet periodically samples image frames and stores the frames on an SD card.
Without \teedriver{}s, such a trustlet cannot enjoy secure IO path: it has to invoke the OS for the needed drivers. 
With \teedriver{}s, 
the trustlet code simply includes one header file and invokes the two interfaces 
for camera and MMC respectively.
Corresponding to the invocations, the replayer selects two \template{}s: one for image capturing and the other for writing 256 blocks. 
To store each frame which is 1--2 MB, the replayer invokes the latter multiple times. 
We have measured that capturing each frame takes 3.7s, in which most is spent on initializing the camera and storing the image only takes 154 ms. 



\section{Discussions}
\label{sec:discussion}

\paragraph{A simpler recorder without symbolic tracing}
To generate driverlets, the developers rely on symbolic tracing.
While the tool (i.e. DBT) exists for popular hardware such as \rpi{}, it may not be always available or easily accessible.
Under such circumstances, developers may resort to tracing the \textit{concrete} interactions at the three interfaces (\S\ref{sec:bkgnd:model}), exercised by desired record campaigns, 
as has been demonstrated in prior work~\cite{tinystack}. 
Additionally, our experiences in Section~\ref{sec:impl} and \ref{sec:case} show that the generated constraints and taints are likely simple, which entail manageable efforts to verify the state transition paths.

\paragraph{Applicability to TEEs other than TrustZone}
Despite we choose \arm{} TrustZone as a key use case for driverlets, driverlets themselves rely only on secure IO and are not tied to \arm{} TrustZone nor a specific TEE.
For instance, driverlets can be used in Keystone of RISC-V~\cite{keystone}, which achieves secure IO via PMP (for memory/registers isolation) and processor M-mode (for routing IRQs);
for SGX which lacks native secure IO support, driverlets require additional techniques~\cite{sgxio}.
Driverlets are not tied to a specific TEE kernel or host environment, neither.
As long as the host environment implements replayer and replay events correctly, driverlets should work out-of-box, e.g. in TEE kernels such as Trusty~\cite{trusty-web} and QTEE~\cite{qtee-doc} or unikernels~\cite{unikernel}.

\section{Related Work}
\label{sec:related}
\paragraph{Driver reuse}
To reuse drivers, some ``lift and shift''~\cite{rumpkernel,picodriver,sgx-lkl,microdrivers}; 
some trim down simple drivers~\cite{optee};
Our system shares same goal.
Different from them, our system derives drivers by a novel use of record and replay. 
\paragraph{Record and replay} is well-known and primarily applied to bug finding~\cite{rnr-bugfinding, rnr-windows-bugfinding,reVirt,r2} and security analysis~\cite{sanity, replayConfusion, rtag}.
It also enables offloading~\cite{reran,mobiplay}, emulation~\cite{mahimahi}, and cheap versioning~\cite{knockoff}.
We are inspired by them to reproduce a subset of program behaviors, e.g. device/driver interactions. 
The key difference is we generalize replay inputs such that replay completes requests beyond those recorded. 
Quite related to driverlets, GPUReplay~\cite{tinystack} also records and replays the device interactions at hardware/software boundary.
Compared to it, driverlets face a different challenge: how to generalize and parameterize recordings; this challenges necessitates a new construct -- interaction template;
by contrast, GPUReplay does not parameterize input events.



\paragraph{Program analysis techniques }
have been widely applied by extensive works for testing~\cite{symdrive} and finding vulnerabilities~\cite{dr.checker} in kernel drivers and excavating data structures (e.g. in binaries~\cite{howard-data-structure-excavator, binrec} and network protocols~\cite{discoverer}).
Inspired by them, our system is built atop well-known program analysis techniques. 
Differently, we use those techniques for a distinct goal: reuse device state transition paths.
\paragraph{Trusted execution environment}
is commonly used to shield trustlets from untrusted host OS~\cite{trustshadow,rubinov16icse,occlumency,darknetz}.
Lacking storage drivers, the trustlets delegate IO to OS~\cite{sbd,s2net,streamboxtz} and mediate their accesses~\cite{secloak}.
Similar with them, we leverage TEE's strong security guarantees;
differently, we provide key missing device drivers for them.
Some works bring drivers to TEE, e.g. IPU~\cite{rushmore}, GPU~\cite{tinystack}. 
Compared to them, which are point solutions to individual devices, we present a holistic approach to systematically derive a set of drivers.

\section{Conclusions}

We present a novel approach to deriving device drivers for TrustZone.
Our toolkit records driver/device interactions from a gold driver and accordingly distills \template{}s;
by replaying a template with new dynamic inputs, the \teedriver{} completes requests beyond the one being recorded while assuring correctness.
We build the recorder/replayer and show that \teedrivers{} have practical performance on MMC, USB, and VCHIQ.
Driverlets fix the key missing link for secure IO, and for the first time open a door for trustlets to access complex yet essential devices.

\section*{Acknowledgment}

The authors were supported in part by NSF awards \#1846102, \#1919197, and \#2106893.
The authors thank the anonymous reviewers and the shepherd Prof. Marios Kogias for their insightful feedback.



\bibliographystyle{abbrv}




\end{document}